\documentclass[twocolumn,aps,pra,showpacs,superscriptaddress]{revtex4-1}
\usepackage{color}
\usepackage{amsmath}
\usepackage{amssymb}
\usepackage{graphicx}
\usepackage{esint}
\usepackage{soul}
\makeatletter

\usepackage{color}
\newcommand{\old}[1]{}
\newcommand{\comment}[1]{}
\newcommand{\new}[1]{#1}
\usepackage{rotating}\usepackage{braket}\usepackage{amsthm}\usepackage{bbm}

\makeatother

\begin{document}

\title{
Dissipation enabled efficient excitation transfer from a
single photon 
to a single quantum
emitter
}

\author{N. Trautmann}

\affiliation{Institut f\"ur Angewandte Physik, Technische Universit\"at Darmstadt,D-64289,
Germany}

\author{G. Alber}

\affiliation{Institut f\"ur Angewandte Physik, Technische Universit\"at Darmstadt,D-64289,
Germany}
\begin{abstract}
We propose a scheme for triggering a dissipation dominated highly efficient excitation transfer
from a single 
photon wave packet
to a single quantum emitter.  
This single photon induced optical pumping turns dominant
dissipative processes, such as spontaneous photon emission by the emitter or cavity decay, into
valuable tools for quantum information processing and quantum communication.
It works for an arbitrarily shaped single photon wave packet with sufficiently small bandwidth
provided a matching condition is satisfied which balances the dissipative rates involved. 
Our scheme does not require additional laser pulses or quantum feedback
and\old{ is not restricted to
highly mode selective cavity quantum electrodynamical
architectures} \new{does not rely on high finesse optical resonators}.
In particular, it can be used
to enhance significantly the  coupling of a single photon to a single quantum
emitter implanted in a one dimensional waveguide or even in a free space scenario.
We demonstrate the usefulness of our scheme for building a deterministic
quantum memory and a deterministic frequency
converter between photonic qubits of different wavelengths. 
\end{abstract}
\date{\today}

\maketitle

\section{Introduction}
Achieving highly efficient excitation transfer from a single photon 
to \new{a} material quantum system
with the possibility of a controlled manipulation of the resulting quantum state
is a crucial 
prerequisite
for advancing quantum technology with potential applications
ranging from quantum communication \cite{Kimble} and computation
\cite{Nielsen} to fundamental tests of quantum mechanics
\cite{hensen2015experimental}.
Coherent quantum processes
provide powerful tools for such an excitation transfer on the single photon level.

With the help of electromagnetically induced transparency
\cite{fleischhauer2000dark,phillips2001storage}, for example,
a single photon wave packet of
quite arbitrary pulse shape can be stored in a collective excitation of a macroscopically
large number of atoms
\cite{phillips2001storage} or in a solid \cite{de2008solid}. However, the
controlled manipulation of the resulting
macroscopic
excitation for purposes of quantum information processing is highly challenging.
In contrast, excitation transfer from a single photon to a
single quantum emitter, such as a trapped atom,
offers the advantage that the resulting quantum state
can be manipulated with high accuracy
\cite{Cirac1997,Ritter,duan2004scalable,reiserer2013nondestructive,reiserer2014quantum,kalb2015heralded}.
Based on
coherent processes an early protocol
suitable for scalable photonic quantum information processing
has been proposed by Cirac et al.
\cite{Cirac1997} and has been implemented experimentally by
Ritter et al.\cite{Ritter}.
However, this protocol requires detailed knowledge of
shape and of arrival time of the photon wave packet for
triggering an appropriate coherent
laser-induced process.
A coherent scheme overcoming the complications of such a conditional pulse shaping
has been proposed by Duan and Kimble
\cite{duan2004scalable}. 
It takes advantage of a trapped atom's state dependent frequency shift of the cavity mode 
which results
in a phase flip of an incoming single photon
reflected by the cavity.
This scheme has been used to
build a nondestructive photon detector \cite{reiserer2013nondestructive},
a quantum gate between a matter and a photonic qubit \cite{reiserer2014quantum},
and a quantum memory for the heralded storage of a single photonic
qubit \cite{kalb2015heralded}. However, for the heralded storage
of a single photonic qubit
the outgoing photon has to be measured
and quantum feedback has to be applied. 
Thus, the efficiency is limited by the efficiency of the single photon detector.
\begin{figure}[t]
\begin{center}
\includegraphics[width=8.5cm]{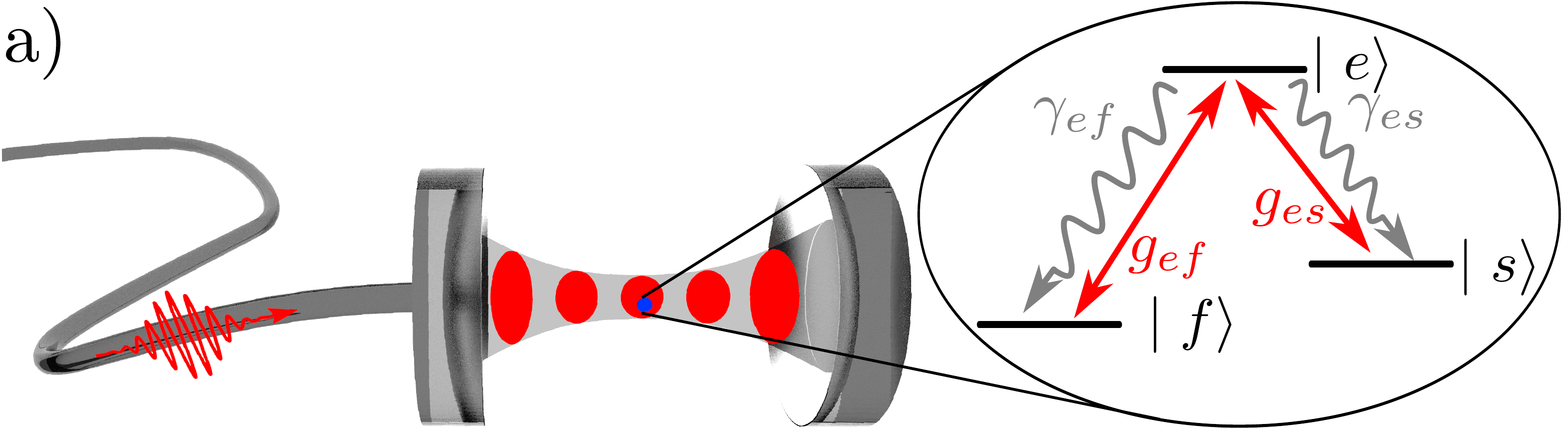}\\
\includegraphics[width=8.5cm]{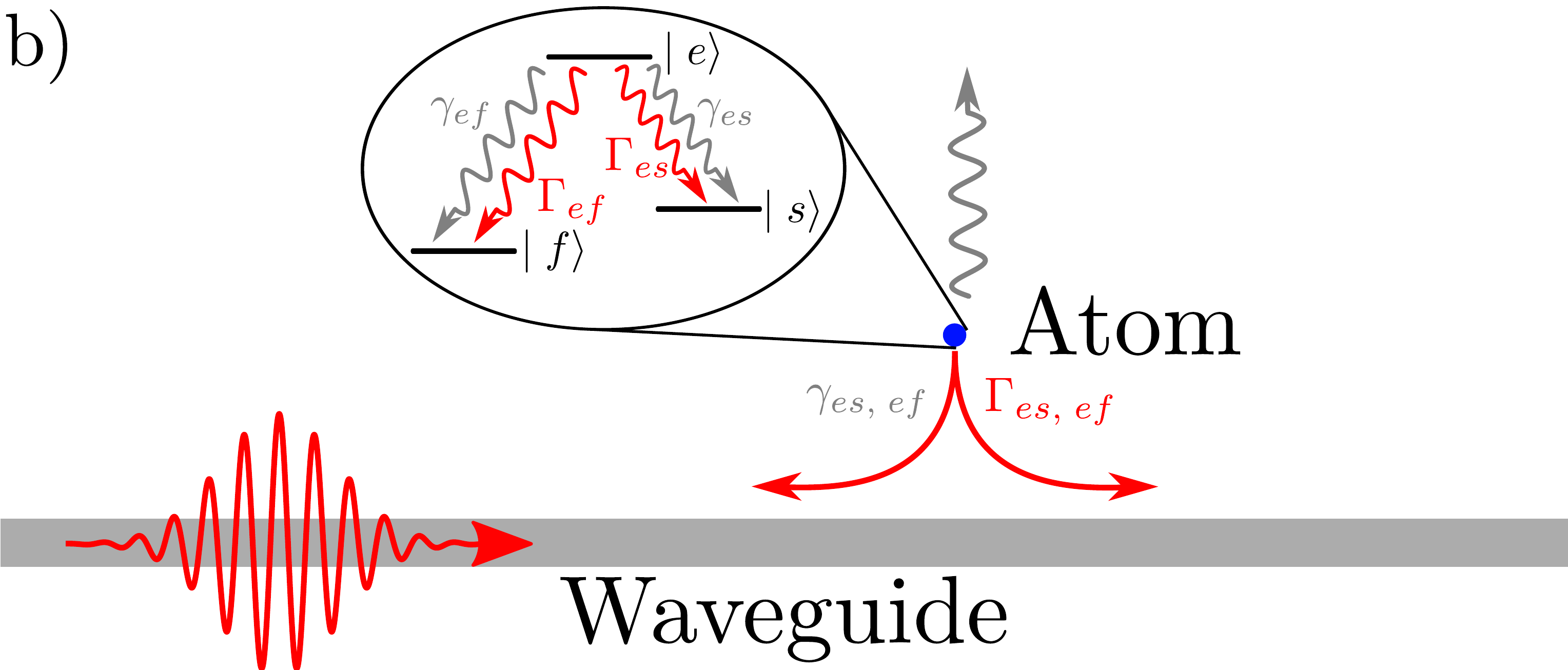}
\par\end{center}
\caption{\new{(a):} Schematic representation of a fiber \new{and cavity} based scenario\old{ with strong mode selection by an optical resonator}.
A single three level atom is trapped inside the cavity and a single photon is propagating through the fiber.
\new{(b): Schematic representation of a three level atom coupled to the evanescent field surrounding a one dimensional waveguide with a single photon propagating along this waveguide.}
 \label{fig:Setup}}
\end{figure}

The natural question arises whether 
it is possible
to achieve highly efficient
excitation transfer
from a single 
photon with a rather arbitrary pulse shape
to a single quantum emitter also in a way
that 
the challenging complications arising from conditional tailoring of
laser pulses and from
imperfections affecting
postselective photon detection processes can be circumvented.
We present such a scheme which is capable of accomplishing
basic tasks of quantum information processing, such as implementing
a deterministic single atom quantum memory or a deterministic
frequency converter for a photonic qubit. Contrary to previous proposals based on coherent quantum processes 
our scheme is enabled by an appropriate balancing of dissipative processes,
such as spontaneous photon emission and cavity decay.
It is demonstrated that this way
a single photon wave packet
of rather arbitrary shape can trigger a highly efficient excitation transfer to a material quantum emitter.
For photon wave packets with sufficiently small bandwidths
the high efficiency
of this excitation transfer is independent of the photon wave packet's shape.

This single photon induced optical pumping \cite{happer1972optical}
does not require \old{mode selection by} an optical resonator \old{\cite{notion_modeselection}}
and is applicable to various scenarios including
highly efficient coupling of a single atom to a single photon propagating in a one dimensional waveguide, such as a nanowire \cite{akimov2007generation,schuller2010plasmonics},
or a nanofiber \cite{vetsch2010optical}, or in a coplanar waveguide (circuit QED) \cite{you2011atomic}, or
even in free space \cite{maiwald2012collecting, fischer2014efficient}.
\new{A schematic representation of a suitable cavity and fiber based scenario as well as a schematic representation of an
atom coupled to the evanescent field surrounding a one dimensional waveguide is depicted in Fig. \ref{fig:Setup} (a) and (b).}

\new{The scheme presented in this article paves the way to scalable quantum communication networks, 
as it relaxes the requirements on the synchronization of the nodes of the network, i.e. detailed knowledge on the arrival time and shape of the photons is not required,
and it is not limited by the efficiency of single photon detectors.}

\new{The body of this article is divided into four parts. In Sec. \ref{Sec_model} we introduce the Hamiltonians for modeling the dynamics for a fiber and cavity based scenario as well as for a waveguide or free space scenario.
Based on these models, we analyze the dynamics of these systems in Sec. \ref{Sec_Dynamics}, and derive the conditions for triggering an efficient excitation transfer.
A key step in the derivation of these analytical results is an adiabatic approximation. In Sec. \ref{Sec_Numerics}, we supplement these analytical results by a numerical investigation.
We show that our scheme allows to trigger an efficient state transfer also with photons of a finite bandwidth.
Finally, in Sec. \ref{Sec_Applications} we  show possible application of our scheme for building a deterministic
quantum memory and a deterministic frequency
converter between photonic qubits of different wavelengths.} 

\section{Quantum optical model}\label{Sec_model}
Let us start by considering a fiber and cavity based system
as schematically depicted in Fig. \ref{fig:Setup} \new{(a)}. 
A three level atom 
is interacting resonantly with two
modes of a surrounding high finesse cavity. A photon propagating
through a fiber can enter this cavity by transmission through
a mirror of the single-sided cavity.
Spontaneous decay of the three level atom is modeled by coupling to the
continua of electromagnetic field modes orthogonal to the modes of the
resonant cavity and of the fiber.
We assume that the dipole
and rotating wave approximations are applicable. In the interaction
picture the Hamiltonian reads 
\begin{small}
\begin{eqnarray}
\hat{H}_{\text{int}}^{\new{\text{cavity}}}(t)/\hbar & = & \left[i g_{es}\hat{a}_{es}^{\text{C}\dagger}\mid s\rangle\langle e\mid+i g_{ef}\hat{a}_{ef}^{\text{C}\dagger}\mid f\rangle\langle e\mid+\text{H.c.}\right]\nonumber\\
 &  & +\left[\sqrt{2\kappa_{es}}\hat{a}^{\text{F}_{es}}(t)\hat{a}_{es}^{\text{C}\dagger}+\text{H.c.}\right]\nonumber\\
 &  & +\left[\sqrt{2\kappa_{ef}}\hat{a}^{\text{F}_{ef}}(t)\hat{a}_{ef}^{\text{C}\dagger}+\text{H.c.}\right]\nonumber\\
 &  & -\frac{1}{\hbar}\left[e^{-i\omega_{es}\left(t-t_{0}\right)}\mathbf{d}_{es}\cdot\hat{\mathbf{E}}_{\text{B}_{es}}^{-}(t)\mid s\rangle\langle e\mid+\text{H.c.}\right]\nonumber\\
 &  & -\frac{1}{\hbar}\left[e^{-i\omega_{ef}\left(t-t_{0}\right)}\mathbf{d}_{ef}\cdot\hat{\mathbf{E}}_{\text{B}_{ef}}^{-}(t)\mid f\rangle\langle e\mid+\text{H.c.}\right],\nonumber\\
 \label{H_cavity}
\end{eqnarray}
\end{small}
$\hat{a}_{es}^{\text{C}}$ and $\hat{a}_{ef}^{\text{C}}$ being the annihilation operators of the cavity modes.
These modes couple
resonantly to the
atomic transitions $\mid e\rangle\leftrightarrow\mid s\rangle$ and
$\mid e\rangle\leftrightarrow\mid f\rangle$ with the atomic transition
frequencies $\omega_{es}$ and $\omega_{ef}$ and the corresponding vacuum Rabi-frequencies $ g_{es}$ and $ g_{ef}$.
These couplings are either due to different polarizations or different frequencies of the cavity
modes. \new{Strictly speaking, the modes in the cavity are not modes but isolated resonances, i.e. bound states in the continuum, as they have a finite spectral width, which is determined by the corresponding cavity loss rates $2\kappa_{es}$ and $2\kappa_{ef}$.
In the following we assume that the photons in the cavity dominantly leak out through one mirror of the single-sided cavity directly into the fiber.
We take this into account by using a Fano-Anderson-type model \cite{fano1935sullo,fano1961effects,anderson1961localized}. Hereby,
}
the coupling
between cavity modes and fiber modes can be described by collective annihilation operators
of the fiber \cite{gardiner2004quantum}, i.e.
\begin{eqnarray*}
\sqrt{2\kappa_{k}}\hat{a}^{\text{F}_{k}}(t) & = & \sum_{j\in \text{F}_{k}}c_{j}\hat{a}_{j}e^{-i\left(\omega_{j}-\omega_{k}\right)(t-t_{0})}\text{ for }k\in\left\{ es,ef \right\}\\
\end{eqnarray*}
with $\hat{a}_{j}~(j\in \text{F}_{es} (\text{F}_{ef}))$ describing
the orthogonal fiber modes with frequencies $\omega_{j}$ \new{coupling to the cavity mode described by the annihilation operator $\hat{a}_{es}^{\text{C}}$  ($\hat{a}_{ef}^{\text{C}}$).}\old{ coupling to the atomic transitions
$\mid e\rangle\leftrightarrow \mid s\rangle$ 
($\mid e\rangle\leftrightarrow~ \mid f\rangle$)\new{.}}\old{
and with
$2\kappa_{es}$ ($2\kappa_{ef}$) denoting the associated
cavity loss rates due to
photon exchange between the cavity and the fiber modes $\text{F}_{es}$ ($\text{F}_{ef}$)
through one of the  mirrors.} The coupling of the atomic transition $\mid e\rangle\leftrightarrow\mid s\rangle$ 
($\mid e\rangle\leftrightarrow\mid f\rangle$) to the 
electromagnetic background modes $\text{B}_{es}$ ($\text{B}_{ef}$)
is characterized by the electric field operator whose negative frequency parts is denoted
$\hat{\mathbf{E}}_{\text{B}_{es}}^{-}(t)$ ($\hat{\mathbf{E}}_{\text{B}_{ef}}^{-}(t)$) and the dipole
matrix element $\mathbf{d}_{es}$  ($\mathbf{d}_{ef}$).

\new{As it turns out, our scheme can also be applied in the absence of a cavity. It can be used to couple a photon propagating along a waveguide to a quantum emitter placed in the vicinity of a waveguide as depicted in Fig. \ref{fig:Setup} (b).
This can even be generalized to coupling a photon propagating in free space to a single atom or ion.
In the interaction picture, the Hamiltonian describing the dynamics of a three level atom coupling to a photon propagating along a one dimensional waveguide or in free space is of a similar form and reads
\begin{eqnarray*}
\hat{H}_{\text{int}}^{\text{wg}}(t) & = & -\left[e^{-i\omega_{es}\left(t-t_{0}\right)}\mathbf{d}_{es}\cdot\mathbf{\hat{E}}^{-}(\mathbf{x}_{A},t)\mid s\rangle\langle e\mid+\text{H.c.}\right]\\
 &  & -\left[e^{-i\omega_{ef}\left(t-t_{0}\right)}\mathbf{d}_{ef}\cdot\mathbf{E}^{-}(\mathbf{x}_{A},t)\mid f\rangle\langle e\mid+\text{H.c.}\right]\,.
\end{eqnarray*}
Hereby, $\hat{\mathbf{E}}^{-}(\mathbf{x}_{A},t)$ and $\hat{\mathbf{E}}^{+}(\mathbf{x}_{A},t)$ are the negative and positive frequency parts of the electric field operator. The detailed description of the waveguide or the free space scenario at hand
is encoded in the structure of the modes entering the field operator $\hat{\mathbf{E}}^{\pm}(\mathbf{x}_{A},t)$.
For analyzing the waveguide scenario we will assume that we can split the set of modes of
the electromagnetic radiation field into four subsets of orthogonal mode functions, i.e. solutions of the Helmholtz equation with appropriate boundary conditions, corresponding
to the four photonic reservoirs $\text{F}_{es}$, $\text{F}_{ef}$, $\text{B}_{es}$,
and $\text{B}_{ef}$. The reservoirs $\text{F}_{es}$ and $\text{F}_{ef}$
involve the modes describing the propagation of photons along
the waveguide, with the reservoir $\text{F}_{es}$ coupling to the
transition $\mid e\rangle\leftrightarrow\mid s\rangle$ and with the reservoir
$\text{F}_{ef}$ coupling to the transition $\mid e\rangle\leftrightarrow\mid f\rangle$.
The reservoirs $\text{B}_{es}$ and $\text{B}_{ef}$ correspond to
the modes describing the propagation of photons not guided by the
waveguide. They are used to model the emission of photons out of the waveguide
and are also grouped according to their coupling to the transitions
$\mid e\rangle\leftrightarrow\mid s\rangle$ and $\mid e\rangle\leftrightarrow\mid f\rangle$.
Accordingly, we can decompose the electric field operator
\[
\mathbf{\hat{E}}^{\pm}(\mathbf{x},t)=\mathbf{\hat{E}}_{\text{F}_{es}}^{\pm}(\mathbf{x},t)+\mathbf{\hat{E}}_{\text{F}_{ef}}^{\pm}(\mathbf{x},t)+\mathbf{\hat{E}}_{\text{B}_{es}}^{\pm}(\mathbf{x},t)+\mathbf{\hat{E}}_{\text{B}_{ef}}^{\pm}(\mathbf{x},t)
\]
into four parts corresponding to these four reservoirs. In general, the splitting of the set of modes into the four reservoirs listed above is connected with some approximations, as effects such as the damping of photons propagating along the waveguide are not described by this ansatz.
However, our model allows us to take the most important loss effect, the emission of a photon by the atom out of the waveguide into account. Furthermore, a more detailed model, transcending the splitting of the set of modes into the four reservoirs
requires detailed knowledge of the structure of the mode functions and, hence, depends on the details of the experimental setup under consideration. As we intend to discuss general waveguide scenarios our subsequent discussion is based on the model introduced above
which allows us to take the most important physical effects into account.
} 

\section{Dynamics and conditions for an efficient excitation transfer}\label{Sec_Dynamics}
\new{In this section we investigate the dynamics of the quantum optical model of Sec. \ref{Sec_model}. We derive a set of conditions for triggering an efficient state transfer of the atom by a single incoming photon.}
\subsection{Cavity}
\begin{figure}[t]
\begin{center}
\includegraphics[width=8.5cm]{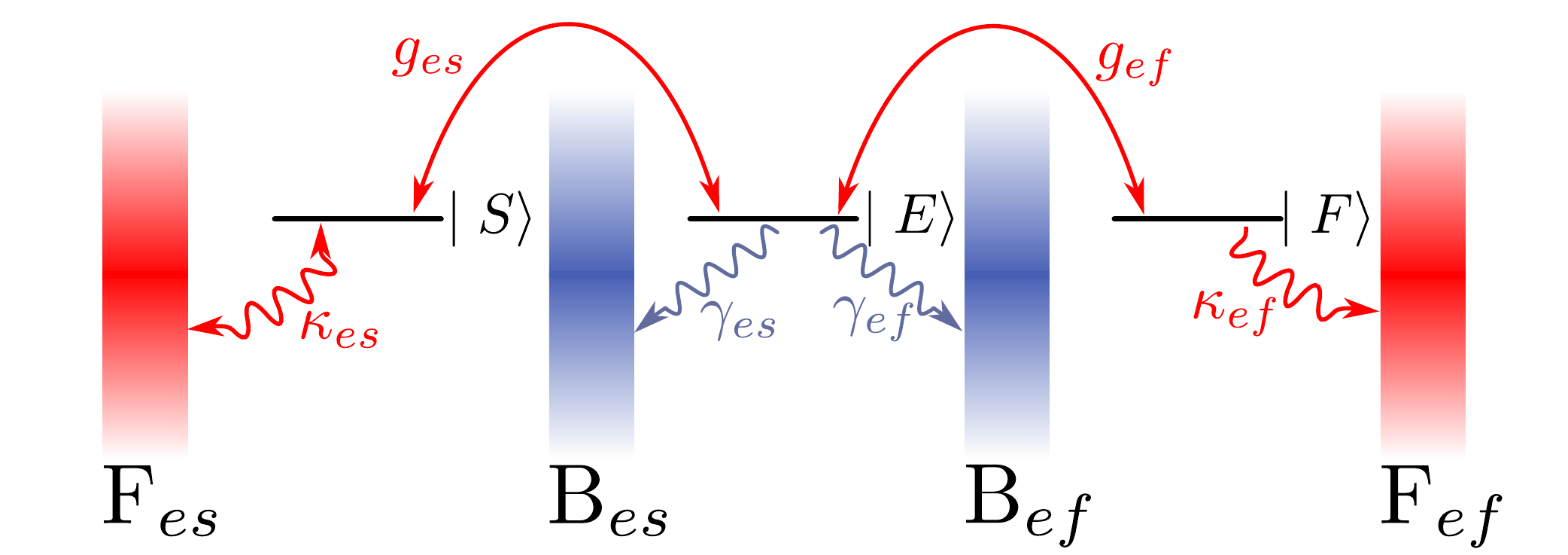}
\par\end{center}

\caption{\new{Schematic representation of the couplings between the states $\mid E\rangle$, $\mid S\rangle$ and $\mid F\rangle$ as well as their couplings to the reservoirs $\text{F}_{es}$, $\text{F}_{ef}$, $\text{B}_{es}$, and $\text{B}_{ef}$.}
 \label{fig:coupling_scheme}}
\end{figure}

\new{We start with the cavity and fiber based scenario described by the Hamiltonian $\hat{H}_{\text{int}}^{\text{cavity}}(t)$ of Eq. (\ref{H_cavity}).}
We consider an initial state $\mid\psi(t_{0})\rangle$ in which a single photon  with frequencies
centered around $\omega_{es}$ is propagating though the fiber towards the left mirror. The
remaining parts of the radiation field are assumed to be in the vacuum state
and the atom is initially prepared in state $\mid s\rangle$, i.e.
\[
\mid\psi(t_{0})\rangle=\mid s\rangle
\mid\psi_{\text{in}}\rangle^{\text{F}_{es}}
\mid 0\rangle^{\text{F}_{ef}}
\mid 0\rangle_{es}^{\text{C}}
\mid0\rangle_{ef}^{\text{C}}
\mid0\rangle^{\text{B}_{es}}
\mid0\rangle^{\text{B}_{ef}}
.
\]
The initial
state of the single photon propagating through the fiber
is denoted
$\mid\psi_{\text{in}}\rangle^{\text{F}_{es}}$ and
$\mid0\rangle_{es}^{\text{C}}$, $\mid0\rangle_{ef}^{\text{C}}$,
$\mid0\rangle^{\text{F}_{ef}}$
$\mid0\rangle^{\text{B}_{es}}$ $\mid0\rangle^{\text{B}_{ef}}$ are the vacuum states of the cavity modes,
of the initially unoccupied fiber modes 
and of the modes of the electromagnetic background. By applying the methods developed
in \cite{mollow1975pure,comment_Fano,gardiner2004quantum} the dynamics of the pure
quantum state $\mid\psi_{\text{P}}(t)\rangle$ can be described by the equation
\begin{eqnarray}
\frac{d}{dt}\mid\psi_{\text{P}}(t)\rangle  = \new{-i} \hat{G}\mid\psi_{\text{P}}(t)\rangle+\mid S\rangle\sqrt{2\kappa_{es}}f_{\text{in}}(t)\;
\label{equationofmotion}
\end{eqnarray}
with the non-Hermitian generator 
\begin{eqnarray}
\hat{G} & = & \new{i}\left[ g_{es}\mid S\rangle\langle E\mid +  g_{ef}\mid F\rangle\langle E\mid-\text{H.c.}\right]
-\\
&&\new{i}\kappa_{es}\mid S\rangle\langle S\mid-
\new{i}\kappa_{ef}\mid F\rangle\langle F\mid-
\new{i}\frac{\gamma_{ef}+\gamma_{es}}{2}\mid E\rangle\langle E\mid.\nonumber
\end{eqnarray}
\new{The anti-Hermitian part of $\hat{G}$ describes the depletion of the population out of the subspace spanned by $\ket{S}$, $\ket{E}$, $\ket{F}$.}
The atomic and photonic excitations inside the cavity are described by the
orthonormal quantum states
\begin{eqnarray*}
\mid E\rangle & \equiv & \mid e\rangle
\mid0\rangle^{\text{F}_{es}}\mid0\rangle^{\text{F}_{ef}}
\mid0\rangle_{es}^{\text{C}}\mid0\rangle_{ef}^{\text{C}}
\mid0\rangle^{\text{B}_{es}}\mid0\rangle^{\text{B}_{ef}},\\
\mid S\rangle & \equiv & \mid s\rangle
\mid0\rangle^{\text{F}_{es}}\mid0\rangle^{\text{F}_{ef}}
\mid1\rangle_{es}^{\text{C}}
\mid0\rangle_{ef}^{\text{C}}
\mid0\rangle^{\text{B}_{es}}\mid0\rangle^{\text{B}_{ef}},\\
\mid F\rangle & \equiv & \mid f\rangle
\mid0\rangle^{\text{F}_{es}}\mid0\rangle^{\text{F}_{ef}}
\mid0\rangle_{es}^{\text{C}}
\mid1\rangle_{ef}^{\text{C}}
\mid0\rangle^{\text{B}_{es}}\mid0\rangle^{\text{B}_{ef}}.
\end{eqnarray*}
The inhomogeneity of Eq. (\ref{equationofmotion})
with amplitude
\begin{eqnarray}
i f_{\text{in}}(t)&=&{}^{\text{F}_{es}}\langle0\mid
\hat{a}^{\text{F}_{es}}(t)\mid\psi_{\text{in}}\rangle^{\text{F}_{es}}
\end{eqnarray}
characterizes the
incoming single photon.
The
spontaneous decay rates of the dipole transitions $\mid e\rangle\leftrightarrow\mid s\rangle$
and $\mid e\rangle\leftrightarrow\mid f\rangle$ are denoted
$\gamma_{es}$ and $\gamma_{ef}$.
\new{A schematic representation  of the coupling of the states $\mid E\rangle$, $\mid S\rangle$ and $\mid F\rangle$ among each other as well as their couplings to the reservoirs $\text{F}_{es}$, $\text{F}_{ef}$, $\text{B}_{es}$, and $\text{B}_{ef}$ as described by
Eq. (\ref{equationofmotion}) is illustrated in Fig. \ref{fig:coupling_scheme}.}
\new{In appendix \ref{appendix}, we derive an equation similar to  Eq. (\ref{equationofmotion}) for the waveguide scenario in the absence of a cavity.
The derivation of Eq. (\ref{equationofmotion}) follows the same lines. }

The solution of Eq. (\ref{equationofmotion}) is given by 
\begin{eqnarray}
\mid\psi_{\text{P}}(t)\rangle=\sqrt{2\kappa_{es}}\int_{t_{0}}^{t}e^{\new{-i}\hat{G}(t^{\prime}-t_{0})}\mid S\rangle f_{\text{in}}(t^{\prime})dt^{\prime}\,.
\end{eqnarray}
We concentrate on the adiabatic dynamical regime in which the bandwidth of the incoming single photon wave packet, i.e.
\begin{eqnarray}
\Delta\omega=\sqrt{\int_{\mathbb{R}}\mid\frac{d}{dt}f_{\text{in}}(t)\mid^{2}dt/\int_{\mathbb{R}}\mid f_{\text{in}}(t)\mid^{2}dt}\,,
\end{eqnarray}
is much smaller than the eigenfrequencies of the generator $\hat{G}$, i.e.
\begin{eqnarray}
&&\Delta \omega \ll \kappa_{es},~
\left|
\frac{\kappa_{es} + (\gamma_{es}+\gamma_{ef})/2}{2}
\pm
\right. \label{eq:badnwidth}\\
&&
\left.
\left[
\left(\frac{
\kappa_{es} - (\gamma_{es}+\gamma_{ef})/2}{2}
\right)^2
-\mid  g_{es}\mid^2 -\mid  g_{ef}\mid^2
\right]^{1/2}\right|\nonumber.
\end{eqnarray}
Such small bandwidth photons can be produced by the method introduced in \cite{Cirac1997} and implemented in \cite{kuhn2002deterministic,Ritter}.
In this dynamical regime \cite{bender1999advanced}, we arrive  at the approximate result
\begin{eqnarray}\frac{\mid\psi_{\text{P}}(t)\rangle }{\sqrt{2\kappa_{es}}} &=&  f_{\text{in}}(t)\int_{0}^{\infty}e^{\new{-i}\hat{G}t^{\prime}}\mid S\rangle dt^{\prime}=  \new{-i} f_{\text{in}}(t)\hat{G}^{-1}\mid S\rangle\nonumber
\end{eqnarray}
if the initial state has
been prepared long before the wave packet arrives at the cavity,
i.e. $t_{0}\rightarrow-\infty$.
Long after the photon has left the cavity again, i.e. for $t\to~\infty$
the atomic transition probability $P_{\mid s\rangle\rightarrow\mid f\rangle}$ 
between initial and final states $\mid s\rangle$
and $\mid f\rangle$ 
is given by
\begin{eqnarray}
P_{\mid s\rangle\rightarrow\mid f\rangle} & = & \int_{\mathbb{R}}\left[2\kappa_{ef}\mid\langle F\mid\psi_{\text{P}}(t^{\prime})\rangle\mid^{2}\right.
\label{eq:P_cavity} \\
&  & 
 \left.+\gamma_{ef}\mid\langle E\mid\psi_{\text{P}}(t^{\prime})\rangle\mid^{2}\right]dt^{\prime}
  = \eta\int_{\mathbb{R}}\mid f_{\text{in}}(t)\mid^{2}dt\nonumber
\end{eqnarray}
with the efficiency
\begin{eqnarray}
\eta & = & \frac{4\chi_{es}\chi_{ef}}{\left(\chi_{es}+\chi_{ef}\right)^{2}}\frac{2 C_{es}}{1+2 C_{es}}\,,\label{eq:efficiency}
\end{eqnarray}
the transition rates 
 $\chi_{k}  =  \gamma_{k}\left(1+2 C_{k}\right)$,
and the cooperativity parameters $C_{k}=\left|g_{k}\right|^2/(\kappa_{k}\gamma_{k})$ for  $k\in\left\{ es,ef\right\} $.
Hereby, $\int_{t_0}^{t}\mid f_{\text{in}}(t')\mid^{2}dt'$ is the
probability that in the time interval $[t_0,t]$ the single photon 
has arrived at the left mirror (not necessarily entering the cavity).
Provided the photon has arrived at the left mirror (during the time interval $[t_0,\infty)$)
the probability of the resulting excitation
transfer to state $\mid f\rangle$ equals the efficiency $\eta$. For an efficiency close to unity it is required that
\begin{eqnarray}
\chi_{es} & = & \chi_{ef},~
1 \ll   C_{es}.
\label{eq:condition}
\end{eqnarray}
The equality of the transition rates may be viewed as an optical impedance
matching condition. The second requirement implies that for unit efficiency the atom should not
decay from state $\mid e\rangle$ back to state $\mid s\rangle$
by photon emission into the electromagnetic background. Interestingly,
the optimal efficiency achievable is limited by the spontaneous decay
$\mid e\rangle\rightarrow\mid s\rangle$ only and not by photon emission 
into the background modes coupling to the transition $\mid e\rangle \leftrightarrow\mid f\rangle$.
If the spontaneous decay rate $\gamma_{ef}$ is sufficiently large
we do not even require any coupling of the transition $\mid e\rangle\leftrightarrow\mid f\rangle$
to one of the cavity modes in order to achieve unit efficiency. 
Realistic parameters for optical cavities \cite{reiserer2013nondestructive,reiserer2014quantum,kalb2015heralded} result in efficiencies of roughly $92\%$ provided the impedance matching condition is fulfilled.
The high efficiency of the scheme 
can be explained by a destructive interference of the photons getting reflected by the cavity and the photons
which couple into the cavity, interact with the atom and leak out back to the reservoir $F_{es}$.

\new{It might be of interest for experimental implementations to take the  background induced radiative decay of the excited state $\mid e\rangle$ to states other than  $\mid s\rangle$ and $\mid f\rangle$ into account.
By doing so, we obtain the following efficiency for triggering a state transfer (i.e. not finding the atom in the state $\mid s\rangle$ after $t\rightarrow\infty$),
\begin{eqnarray}
\eta & = & \frac{4\chi_{es}\left(\chi_{ef}+\gamma_{eo}\right)}{\left(\chi_{es}+\chi_{ef}+\gamma_{eo}\right)^{2}}\frac{2 C_{es}}{1+2 C_{es}}
\end{eqnarray}
with $\gamma_{eo}$ being the spontaneous decay rate of the state $\mid e\rangle$ to states other than $\mid s\rangle$ and $\mid f\rangle$.
Hence, the rate $\chi_{ef}$ is effectively replaced by $\chi_{ef}+\gamma_{eo}$.
The efficiency for triggering a state transfer and finding the atom finally in state $\mid f\rangle$ is given by
\begin{eqnarray}
\eta_{\mid s\rangle \rightarrow \mid f\rangle} & = & \eta \frac{\chi_{ef}}{\chi_{ef}+\gamma_{eo}}\,.
\end{eqnarray}}
\subsection{Waveguide and free space}
\new{Optimizing this excitation transfer by balancing the relevant dissipation
induced rates as described by condition (\ref{eq:condition})
is not only applicable to
fiber and cavity based scenarios.
Our scheme can also be applied to couple a quantum emitter to a single photon propagating
through a one dimensional waveguide or even in free space. 
 In the following
we assume that initially a single photon resonantly coupling to the
atomic transition $\mid e\rangle\leftrightarrow\mid s\rangle$ is
propagating along a waveguide. Furthermore, the radiative background as
well as the modes of the reservoir $\text{F}_{ef}$ are assumed to be initially
in the vacuum state with the atom being initially prepared in the state $\mid s\rangle$.
Thus, the pure initial state is give by 
\[
\mid\psi(t_{0})\rangle=\mid s\rangle\mid\psi_{\text{in}}\rangle^{\text{F}_{es}}\mid0\rangle^{\text{F}_{ef}}\mid0\rangle^{\text{B}_{es}}\mid0\rangle^{\text{B}_{ef}}
\]
with $\mid\psi_{\text{in}}\rangle^{\text{F}_{es}}$ being the initial
state of the single photon propagating through the waveguide.
As the number
of excitations is a conserved quantity, in the rotating wave approximation the time evolution
of the quantum state of the system is of the form
\begin{eqnarray}
\mid\psi(t)\rangle & = & \psi_{e}(t)\mid e\rangle\mid0\rangle^{\text{F}_{es}}\mid0\rangle^{\text{F}_{ef}}\mid0\rangle^{\text{B}_{es}}\mid0\rangle^{\text{B}_{ef}}\nonumber\\
 &  & +\mid s\rangle\mid\psi_{es}(t)\rangle^{\text{F}_{es},\text{B}_{es}}\mid0\rangle^{\text{F}_{ef}}\mid0\rangle^{\text{B}_{ef}}\nonumber\\
 &&+\mid f\rangle\mid\psi_{ef}(t)\rangle^{\text{F}_{ef},\text{B}_{ef}}\mid0\rangle^{\text{F}_{es}}\mid0\rangle^{\text{B}_{es}}\,
\label{ansatz_wavefunction}
 \end{eqnarray}
 with $\psi_{e}(t)$ being the probability amplitude of finding the
atom in the excited state and with the (unnormalized) state 
$\mid\psi_{es}(t)\rangle^{\text{F}_{es},\text{B}_{es}}$
($\mid\psi_{ef}(t)\rangle^{\text{F}_{ef},\text{B}_{ef}}$) describing
a single photon in the reservoir $\text{F}_{es}$, $\text{B}_{es}$
($\text{F}_{ef}$, $\text{B}_{ef}$).
We can derive the following differential equation characterizing the probability amplitude of finding the atom in an excited state
\begin{eqnarray}
\frac{d}{dt}\psi_{e}(t)&=&-\frac{1}{2}\left(\Gamma_{es}+\Gamma_{ef}+\gamma_{es}+\gamma_{ef}\right)\psi_{e}(t)\nonumber\\
&&+i\sqrt{\Gamma_{es}}f_{in}(t)\;,\label{psi_e_simple_main}
\end{eqnarray}
with
\begin{equation}
\sqrt{\Gamma_{es}}f_{in}(t)=\frac{1}{\hbar}e^{i\omega_{es}\left(t-t_{0}\right)}\langle0\mid\mathbf{d}_{es}^{*}\cdot\left(\mathbf{E}_{es}^{-}\right(\mathbf{x}_{A},t))^{\dagger}\mid\psi_{\text{in}}\rangle^{\text{F}_{es}}\label{eq:f_in}
\end{equation}
describing the influence of the incoming single photon wave packet.
Hereby,
the relevant matter field
couplings in the absence of a cavity are characterized by the rates
of spontaneous photon exchange through the waveguide caused by the transitions $\mid e\rangle\leftrightarrow\mid s\rangle$
and $\mid e\rangle\leftrightarrow\mid f\rangle$, say $\Gamma_{es}$ and
$\Gamma_{ef}$, 
and by the analogous rates 
$\gamma_{es}$ and $\gamma_{ef}$ 
of spontaneous
photon emission out of the waveguide into orthogonal modes of the electromagnetic background. 
The derivation of Eq. (\ref{psi_e_simple_main}) can be found in Appendix \ref{appendix}.
Its solution is given by the 
integral representation 
\begin{eqnarray}
\psi_{e}(t)= & i\sqrt{\Gamma_{es}}\int_{t_{0}}^{t}e^{-\left(\Gamma_{es}+\gamma_{es}+\Gamma_{ef}+\gamma_{ef}\right)\left(t-t^{\prime}\right)/2}f_{in}(t^{\prime})\;.\nonumber\\\label{psi}
\end{eqnarray}}

\new{In the following, we again focus on the adiabatic dynamical
regime in which the bandwidth of the incoming single photon wave packet
$\Delta\omega$ is much smaller than the total spontaneous decay rate
of the excited state, i.e., 
\begin{eqnarray}
 & \Delta\omega\ll\left(\Gamma_{es}+\gamma_{es}+\Gamma_{ef}+\gamma_{ef}\right)/2. \label{eq:condition_badwidth_waveguide}
\end{eqnarray}
 In this adiabatic regime the probability amplitude of finding the
atom in the excited state follows the temporal profile of the incoming
single photon wave packet. Thus, after a partial integration we obtain  from Eq. (\ref{psi}) the approximate result
\begin{eqnarray}
 & \psi_{e}(t) & =i\frac{2\sqrt{\Gamma_{es}}}{\Gamma_{es}+\gamma_{es}+\Gamma_{ef}+\gamma_{ef}}f_{in}(t)\;
\end{eqnarray}
if the initial state has been prepared long before the photon wave packet
arrives at the atom, i.e., $t_{0}\rightarrow-\infty$. As discussed in
the case of a cavity, we can use the above result to evaluate the
probability for triggering an efficient excitation transfer from state
$\mid s\rangle$ to state $\mid f\rangle$. Long after the photon
has left the atom again, i.e. for $t\to~\infty$, the corresponding
atomic transition probability is given by 
\begin{eqnarray}
P_{\mid s\rangle\rightarrow\mid f\rangle} & = & \left(\Gamma_{ef}+\gamma_{ef}\right)\int_{\mathbb{R}}\left|\langle e\mid\psi(t^{\prime})\rangle\right|^{2}\\
 & = & \eta\int_{\mathbb{R}}\left|f_{\text{in}}(t^{\prime})\right|^{2}dt^{\prime}\;.
\end{eqnarray}
 It is possible to achieve 
\begin{eqnarray}
\int_{\mathbb{R}}\left|f_{\text{in}}(t^{\prime})\right|^{2}dt^{\prime}=1
\end{eqnarray}
 in a chiral waveguide \cite{petersen2014chiral,lin2013polarization,mitsch2014quantum} or in a non-chiral waveguide if one side of
the waveguide is terminated by a mirror causing constructive interference
of the electric fields of the incoming and reflected wave packet at
the position of the atom. In general the presence of a mirror, results
in non Markovian effects. However, if the distance of the atom to
the mirror is small compared to $c/\left(\Gamma_{es}+\gamma_{es}+\Gamma_{ef}+\gamma_{ef}\right)$
with $c$ being the speed of light these non Markovian effects can
be neglected. The corresponding efficiency for triggering
the state transfer is given by} 
\begin{eqnarray}
\eta & = & \frac{4\chi_{es}\chi_{ef}}{\left(\chi_{es}+\chi_{ef}\right)^{2}}\frac{\Gamma_{es}}{\Gamma_{es}+\gamma_{es}}\label{eq:efficiency_waveguide}
\end{eqnarray}
\new{with 
\begin{eqnarray}
\chi_{es} & = & \gamma_{es}+\Gamma_{es}\;,\\
\chi_{ef} & = & \gamma_{ef}+\Gamma_{ef}\;
\end{eqnarray}
 in analogy to the transition rates discussed in the fiber and cavity based scenarios.}
\new{Note the close similarity to Eq. (\ref{eq:efficiency}) describing the efficiency for the cavity and fiber based scenario.} 
Optimal state transfer in the absence of a cavity
is achievable 
if 
\begin{eqnarray}
\chi_{es} & = & \chi_{ef},~
\gamma_{es} \ll   \Gamma_{es}.
\label{eq:condition_waveguide}
\end{eqnarray}\old{It can be realized in a
chiral waveguide 
or in a non-chiral waveguide if one side of the waveguide
is terminated by a mirror causing constructive interference of the electric fields of the incoming and reflected wave packet at the position of the atom.}
The emission of photons by the atom into the waveguide is enhanced by confining the field propagating along the waveguide to subwavelength length scales.
For realistic experimental parameters \cite{goban2014atom}, we obtain a transfer efficiency of $32\%$ (provided the impedance matching condition is satisfied).

The free space scenario without any waveguides
can be described by interpreting the modes $\text{F}_{ef}$ as
the only modes which couple to the atomic transition $\mid e\rangle \leftrightarrow\mid f\rangle$ so that $\gamma_{ef} = 0$. In such a case
the continuum $\text{F}_{es}$ may be interpreted as the modes by which the three level system is excited by the incoming single photon with the rate $\Gamma_{es}$.
Consequently, the
background modes $\text{B}_{es}$ have to be interpreted as the additional 
orthogonal background modes to which
the atomic transition $\mid e\rangle \leftrightarrow\mid s\rangle$ can also
decay with rate $\gamma_{es}$. In a free space scenario, perfect excitation of the
transition $\mid s\rangle\to\mid e\rangle$ \cite{stobinska2009perfect}
corresponds to the case
$\gamma_{es} = 0$ in which unit efficiency is achievable for the state transfer
$\mid s\rangle\to\mid f\rangle$ provided the impedance matching condition
$\Gamma_{es} = \Gamma_{ef}$ is fulfilled 
. \new{ The condition $\gamma_{es} = 0$  requires the incoming photon impinging on the atom forming an inward moving dipole wave which couples to the dipole allowed transition $\mid s\rangle\rightarrow\mid e\rangle$ in an optimal way ($\gamma_{es}\ll \Gamma_{es}$ can be realized by using a parabolic mirror \cite{fischer2014efficient}).}
\new{In free space, the impedance matching condition can be fulfilled in two electron atoms with strict LS-coupling, for example. A suitable candidate is ${}^{40}\text{Ca}$.
It has a nuclear spin of $I=0$ and suitable level schemes can be found within the triplet manifolds with the electron spin $S=1$.
A possible level scheme is 
\begin{eqnarray}
 \mid s\rangle&\equiv&\mid3p^{6}3d4s\;^{3}D\; J=3\; m_{J}=1\rangle\,,\nonumber\\
 \mid f\rangle&\equiv&\mid3p^{6}3d4s\;^{3}D\; J=3\; m_{J}=-1\rangle\,,\nonumber\\
 \mid e\rangle&\equiv&\mid3p^{6}3d4p\;^{3}D\; J=3\; m_{J}=0\rangle\,.\nonumber
\end{eqnarray}
This scheme is suitable as the decay from state  $\mid e\rangle$ to $\mid3p^{6}3d4s\;^{3}D\; J=3\; m_{J}=0\rangle$ is not dipole allowed and the decay rates $\Gamma_{es}$ and $\Gamma_{ef}$ are equal.
The only limiting factor stems from the  decay of $\mid e\rangle$ to the manifold $3p^{6}3d4s\;^{3}D\; J=2$. However the decay rate from $\mid e\rangle$ to the manifold $3p^{6}3d4s\;^{3}D\; J=2$ is suppressed by a factor of $8$, as compared to the decay rate to the manifold $3p^{6}3d4s\;^{3}D\; J=3$ \cite{kostlin1964beitrag}.
It might be possible to find more favorable level schemes in other multi-electron atoms or isotopes with a different nuclear spin.
As the decay  of the excited state $\mid e\rangle$ to states other than  $\mid s\rangle$ and $\mid f\rangle$ is a limiting factor, a quantitative description of this effect is of interest.
In close similarity to the fiber and cavity based scenario, we obtain the following efficiency for triggering a state transfer (i.e. not finding the atom in the state $\mid s\rangle$ after $t\rightarrow\infty$),
\begin{eqnarray}
\eta & = & \frac{4\chi_{es}\left(\chi_{ef}+\gamma_{eo}\right)}{\left(\chi_{es}+\chi_{ef}+\gamma_{eo}\right)^{2}}\frac{\Gamma_{es}}{\Gamma_{es}+\gamma_{es}}
\end{eqnarray}
with $\gamma_{eo}$ being the spontaneous decay rate of the state $\mid e\rangle$ into states other than $\mid s\rangle$ and $\mid f\rangle$.
Hence, the rate $\chi_{ef}$ is effectively replaced by $\chi_{ef}+\gamma_{eo}$.
The efficiency for triggering a state transfer and finding the atom finally in state $\mid f\rangle$ is given by
\begin{eqnarray}
\eta_{\mid s\rangle \rightarrow \mid f\rangle} & = & \eta \frac{\chi_{ef}}{\chi_{ef}+\gamma_{eo}}\,.
\end{eqnarray}
Note, that our scheme is surprisingly robust against deviations from the ideal branching ratio. If we consider a level system in which $\Gamma_{es}$ and $\Gamma_{ef}$ differ by a factor of $2$, for example, our scheme still results in a transfer probability of $P_{\mid s\rangle\rightarrow\mid f\rangle}=\frac{8}{9}\approx 89\%$.
In addition, tuning of the spontaneous decay rates may be achieved with the help of additional dressing lasers \cite{marzoli1994laser}, for example.
Thereby, the spontaneous decay rates of the dressed states can be tuned by controlling their overlap with the bare states.}

\section{Numerical investigation}\label{Sec_Numerics}
\begin{figure}
\begin{centering}
\includegraphics[width=8cm]{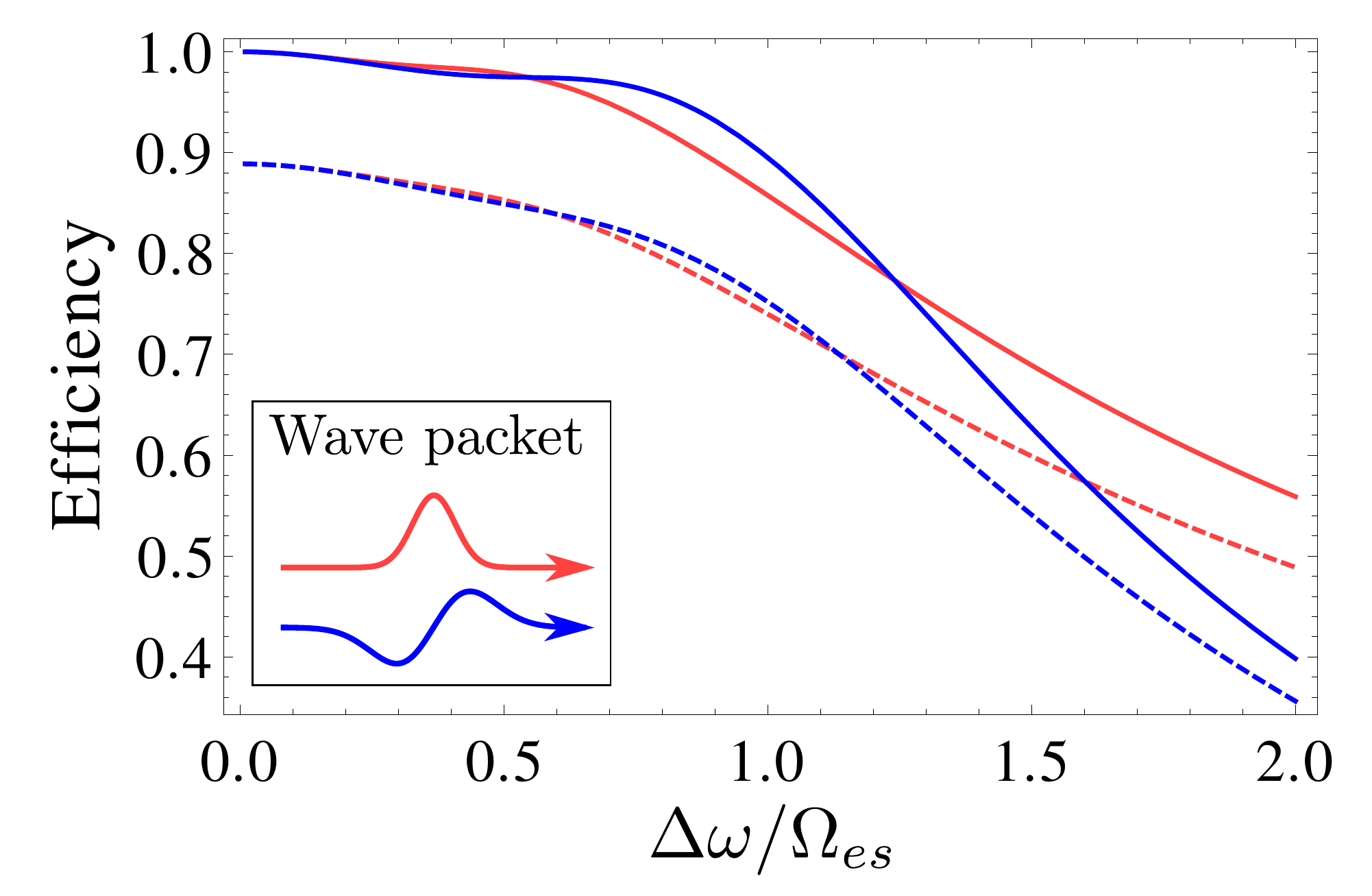}
\par\end{centering}
\caption{Dependence of the state transfer efficiency $\mid s\rangle\rightarrow\mid f\rangle$
on the bandwidth $\Delta\omega$ of the photonic wave packet for the fiber and cavity scenario (in units of $ g_{es}$):
The red (or lighter gray) lines are results of the Gaussian envelope
$f_{\text{in}}^{1}(t)$ and the blue (or darker gray) lines of the antisymmetric envelope $f_{\text{in}}^{2}(t)$ . \new{ The parameters are: $ g_{es}= g_{ef}=\kappa_{es}=\kappa_{ef}$ and
$\gamma_{es},\gamma_{ef}\rightarrow0$ (solid lines; i.e. conditions for high transfer efficiencies in (\ref{eq:condition}) are fulfilled), $ g_{es}= g_{ef}/\sqrt{2}=\kappa_{es}=\kappa_{ef}$ and $\gamma_{es},\gamma_{ef}\rightarrow0$
(dashed lines; $2 \chi_{es}=\chi_{ef}$ first condition in (\ref{eq:condition}) is violated).}
\label{fig:State-transfer-efficiency-cavity}}
\end{figure}

\begin{figure}
\begin{centering}
\includegraphics[width=8cm]{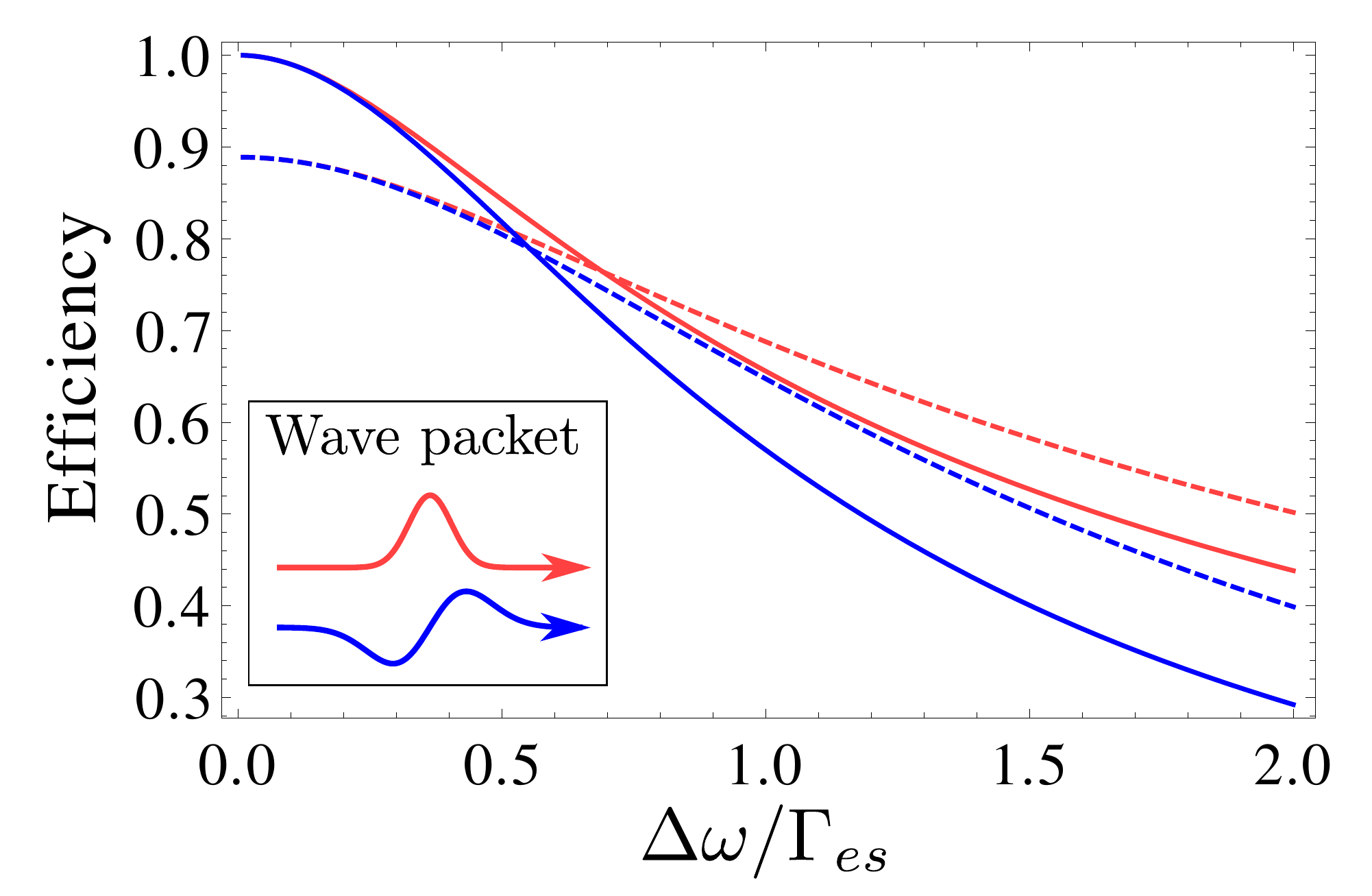}
\par\end{centering}
\caption{\new{Dependence of the state transfer efficiency $\mid s\rangle\rightarrow\mid f\rangle$
on the bandwidth $\Delta\omega$ of the photonic wave packet for the waveguide scenario (in units of $\Gamma_{es}$):
The red (or lighter gray) lines are results of a Gaussian envelope
$f_{\text{in}}^{1}(t)$ and the blue (or darker gray) lines of an antisymmetric envelope $f_{\text{in}}^{2}(t)$ . The parameters are $\Gamma_{es}=\Gamma_{ef}$ and $\gamma_{es}=\gamma_{ef}=0$ (solid lines; i.e. conditions for high transfer efficiencies in (\ref{eq:condition_waveguide}) are fulfilled),
$2\Gamma_{es}=\Gamma_{ef}$ and $\gamma_{es}=\gamma_{ef}=0$ (dashed lines; $2 \chi_{es}=\chi_{ef}$ first condition in (\ref{eq:condition_waveguide}) is violated)}
\label{fig:State-transfer-efficiency-waveguide}}
\end{figure}

For demonstrating the independence of this transition probability
from the shape of the incoming wave packet, we have numerically evaluated the time evolution for two different temporal
envelopes, namely 
a symmetric Gaussian envelope
$$f_{\text{in}}^{(1)}(t)=\sqrt[4]{2\Delta\omega^{2}/\pi}e^{-\Delta\omega^{2}t^{2}}$$
and an antisymmetric envelope
$$f_{\text{in}}^{(2)}(t)=2\sqrt[4]{2/\pi}\Delta\omega^{3/2}te^{-\frac{1}{3}\Delta\omega^{2}t^{2}}/3^{3/4}\,.$$ 
They are normalized so that $\int_{\mathbb{R}}\mid f_{\text{in}}^{(1,2)}(t)\mid^{2}dt=1$,
\new{i.e. the photon certainly 
arrives at the left mirror of the cavity (in the fiber and cavity scenario) or in a suitable waveguide implementation (see previous section) the photon certainly arrives at the atom.
The results are depicted in Fig. \ref{fig:State-transfer-efficiency-cavity} for the cavity scenario and in Fig. \ref{fig:State-transfer-efficiency-waveguide} for the waveguide scenario.}

\new{The solid lines in Fig. \ref{fig:State-transfer-efficiency-cavity} and Fig. \ref{fig:State-transfer-efficiency-waveguide}
correspond to the ideal scenario with high transfer efficiencies as described by Eqs. (\ref{eq:condition}) for the fiber and cavity based scenario and by Eq. (\ref{eq:condition_waveguide}) for the waveguide scenario. As long as the bandwidth of the
incoming photon wave packet is sufficiently
small (see Eqs. (\ref{eq:badnwidth}) and (\ref{eq:condition_badwidth_waveguide})) 
the efficiency of the excitation transfer is close to unity and independent of the shape of the photon wave packet.
The dashed lines in Fig. \ref{fig:State-transfer-efficiency-cavity} and Fig. \ref{fig:State-transfer-efficiency-waveguide}
describe cases with $2\chi_{es}=\chi_{ef}$
so that
a violation of the first condition in Eq. (\ref{eq:condition}) or in Eq. (\ref{eq:condition_waveguide})
limits the efficiency. 
If this impedance matching condition is violated the efficiency for triggering a state transfer is always below unity even in the limit of infinitely small bandwidth photons.}
\old{The spontaneous decay process $\mid e\rangle\rightarrow\mid f\rangle$
does not limit the overall
efficiency and in the extreme case of $ g_{ef} = 0$, for example, 
a nonzero decay rate $\gamma_{ef}$ is even essential for a highly efficient excitation transfer.}

\section{Applications}\label{Sec_Applications}
\new{Our scheme can serve as a basic building block for various tasks of quantum information processing. In this section, we discuss possible applications of our scheme for building a 
single atom single photon quantum memory and for implementing a deterministic frequency converter
of photonic qubits.}

\subsection{Single atom single photon quantum memory}
\begin{figure}
\begin{centering}
\includegraphics[width=8.5cm]{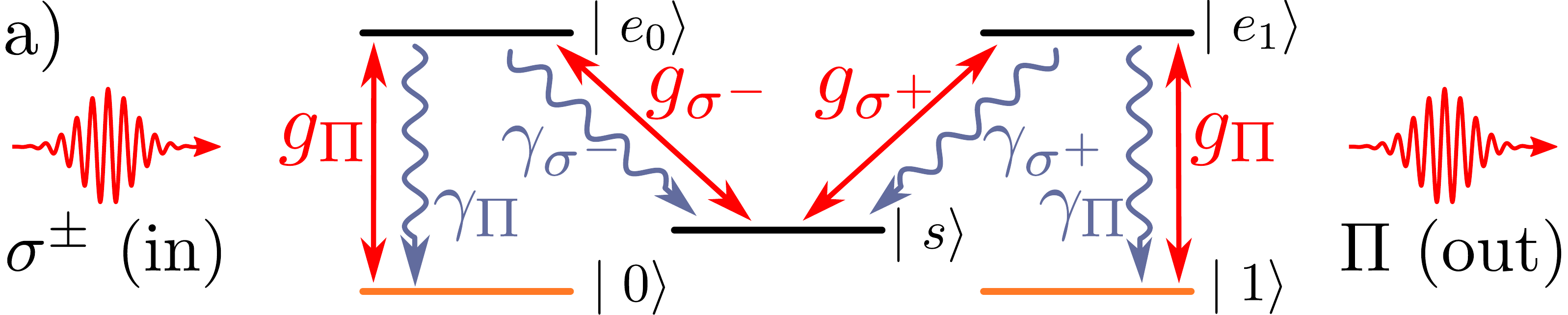}\\
\includegraphics[width=8.5cm]{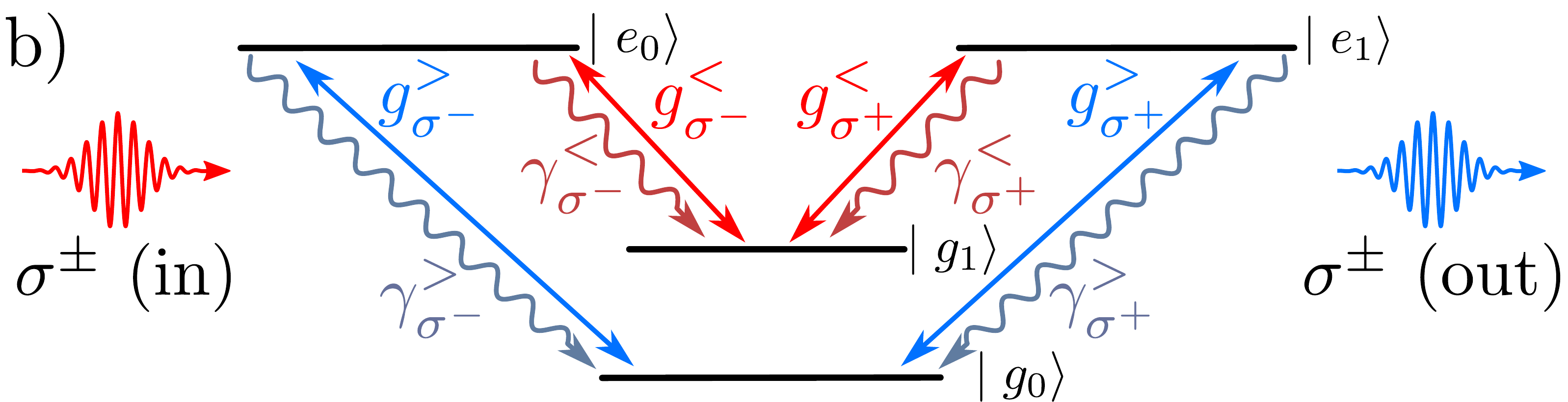}
\par\end{centering}

\caption{An atomic level structure for converting a polarization encoded photonic qubit encoded into a matter qubit (a) and for converting the frequency of a polarization encoded photonic 
qubit (b).\label{fig:storage_conversion}}
\end{figure}

A photonic qubit stored in a polarization degree of
freedom of a photon wave packet,
for example, can be converted to a matter qubit
and stored in the atomic level structure of the atom (for the reverse process see \cite{Cirac1997,Ritter,trautmann2015time}). This can be
achieved by using an atom with a level structure depicted
in Fig. \ref{fig:storage_conversion} (a), for example, with the atom initially prepared in
state $\mid s\rangle$ and with the qubits states $\mid 0\rangle$ and
$\mid 1\rangle$ constituting long lived stable states.
If the properties of the photon emitted during this storage process are
independent of the state of the initial photonic qubit no information
about this photonic input state is transferred to the background or to the fiber modes involved. Thus, the photonic excitation transfer
to the material degrees of freedom
does not suffer from decoherence. For a cavity this condition can be fulfilled by choosing equal vacuum Rabi frequencies
and cavity loss rates for the $\sigma^{\pm}$ transitions,
i.e. $ g_{\sigma^{-}}= g_{\sigma^{+}}$ and $\kappa_{\sigma^{-}}=\kappa_{\sigma^{+}}$.
In the absence of a cavity for this purpose one has to choose equal photon emission
rates into the waveguide, i.e. $\Gamma_{\sigma^{-}}=\Gamma_{\sigma^{+}}$. Hence, the scheme can be used to implement a heralded quantum memory with a fidelity close to unity.
A deterministic quantum memory with near-unit fidelity can be implemented if the impedance matching conditions are fulfilled and $C_{\sigma_{\pm}}\gg 1$ in case of cavity or $\gamma_{\sigma^{\pm}}\ll\Gamma_{\sigma^{\pm}}$ in the absence of a cavity.
Hereby, a coupling of the cavity modes to the $\Pi$ polarized transitions is not required, as these transitions can also be induced by spontaneous decay processes.

\new{A possible level scheme, for the free space scenario can again be found in ${}^{40}\text{Ca}$, for example.
The states $\mid3p^{6}3d4s\;^{3}D\; J=1\; m_{J}=\pm 1\rangle$ could be used to encode the qubit,
the states $\mid3p^{6}3d4p\;^{3}D\; J=1\; m_{J}=\pm 1\rangle$ could serve as intermediate excited states
and the state $\mid3p^{6}3d4s\;^{3}D\; J=1\; m_{J}=0\rangle$ could serve as initial state.
In this level scheme all the branching ratios are equal. The limiting factor is the decay to the manifold $3p^{6}3d4s\;^{3}D\; J=2$.  The decay rate of the states in the manifold $3p^{6}3d4p\;^{3}D\; J=1$
to the manifold $3p^{6}3d4s\;^{3}D\; J=2$ is suppressed by a factor of $3$, as compared to the decay rate to the manifold $3p^{6}3d4s\;^{3}D\; J=1$ \cite{kostlin1964beitrag}.
However, more favorable level schemes might be found in other atoms, isotopes or artificial atoms.
In the waveguide scenario or the cavity scenario the impedance matching condition stated in Eqs. (\ref{eq:condition}) and (\ref{eq:condition_waveguide}) can not 
be connected directly to the dipole matrix elements of the optical transitions, as the modification of the mode structure due to the presence of the waveguide or the cavity (vacuum Rabi frequencies and leakage parameters)
are also of relevance. However, this also allows for a greater tunability of the systems parameters. Hence, in a cavity or a waveguide it might be easier to fulfill the  impedance matching condition than in the free space scenario.}
\subsection{Frequency converter}
Our scheme can also be used for a deterministic frequency converter
of photonic qubits. A possible atomic level structure 
performing 
frequency conversion of a polarization encoded  photonic qubit is depicted in Fig. \ref{fig:storage_conversion} (b).
For converting the frequency of the photon the atom has to
be prepared either in state $\mid g_{0}\rangle$ or in state $\mid g_{1}\rangle$
depending on whether the frequencies of the photon should be decreased or increased. For ensuring
the emission of the resulting photon
into a waveguide
the corresponding vacuum Rabi frequencies or emission rates into the
waveguide have to be sufficiently large. Hence, for a cavity it is required that
$\gamma_{\sigma^{\pm}}^{>}\ll2\mid g_{\sigma^{\pm}}^{>}\mid^{2}/\kappa_{\sigma^{\pm}}^{>}$
and $\gamma_{\sigma^{\pm}}^{<}\ll2\mid g_{\sigma^{\pm}}^{<}\mid^{2}/
\kappa_{\sigma^{\pm}}^{<}$.
In the absence of a cavity, the conditions read $\gamma_{\sigma^{\pm}}^{>}\ll\Gamma_{\sigma^{\pm}}^{>}$
and $\gamma_{\sigma^{\pm}}^{<}\ll\Gamma_{\sigma^{\pm}}^{<}$.
In addition to performing a frequency conversion, the fact that only a single atom is involved allows us to perform 
a nondestructive detection of the photon, by reading out the state of the atom.

\section{Conclusions}
In conclusion, we have proposed a dissipation dominated scheme for triggering highly efficient
excitation transfer from a single photon wave packet
of arbitrary shape but small bandwidth
to a single quantum emitter.
We have shown that by balancing the decay rates characterizing relevant dissipation processes, such as
spontaneous photon emission into waveguides or the electromagnetic background, appropriately these processes
can be turned into a valuable tool for purposes of quantum information processing.
Our scheme offers the advantage that no additional
control of the system by additional laser fields or by postselection is required. 
\new{Thus, the scheme presented in this article paves the way to scalable quantum communication networks
as it relaxes the restrictive requirements on the synchronization of the nodes of the network (detailed knowledge on the arrival time and shape of the photons is not required)
and as it is not limited by the efficiency of single photon detectors.}
We have demonstrated
that our scheme can be applied to a variety of different scenarios including fiber and cavity based architectures as well as architectures without any optical resonators.
It can serve as a basic building block for various protocols relevant
for quantum information processing.
As examples we have discussed setups
for a deterministic single atom single photon quantum memory
and a deterministic frequency converter between photonic qubits of
different wave lengths which could serve as an interface between several
quantum information processing architectures.

\begin{acknowledgments}
  Stimulating discussions with Gerd Leuchs and Peter Zoller are gratefully acknowledged.
  This work is supported by the BMBF Project Q.com, and by the DFG as part of the CRC 1119 CROSSING.
\end{acknowledgments}

\appendix

\section{Derivation for Waveguide}\label{appendix}
\new{
In this appendix, we give a detailed derivation of Eq. (\ref{psi_e_simple_main}), which describes the dynamics of the quantum emitter for the waveguide scenario.
The derivation of Eq. (\ref{equationofmotion}) for the fiber and cavity scenario follows along the same lines. 
We start our considerations with the Ansatz for the time evolution of the wave function of Eq. (\ref{ansatz_wavefunction})
The Schr\"odinger equation induced by the Hamiltonian $\hat{H}_{\text{int}}^{\text{wg}}(t) $ is equivalent to the
following set of differential equations
\begin{widetext}
\begin{eqnarray}
\frac{d}{dt}\psi_{e}(t) & = & \frac{i}{\hbar}\sum_{k\in\{es,ef\}}e^{i\omega_{k}\left(t-t_{0}\right)}\langle0|^{\text{F}_{k}}\langle0|^{\text{B}_{k}}\mathbf{d}_{k}^{*}\cdot\left(\mathbf{\hat{E}}_{\text{F}_{k}}^{+}(\mathbf{x}_{A},t)+\mathbf{\hat{E}}_{\text{B}_{k}}^{+}(\mathbf{x}_{A},t)\right)\mid\psi_{k}(t)\rangle^{\text{F}_{k},\text{B}_{k}}\label{eq:psi_e}\;,\\
\mid\psi_{k}(t)\rangle^{\text{F}_{k},\text{B}_{k}} & = & \frac{i}{\hbar}e^{-i\omega_{k}\left(t-t_{0}\right)}\mathbf{d}_{k}\cdot\left(\mathbf{\hat{E}}_{\text{F}_{k}}^{-}(\mathbf{x}_{A},t)+\mathbf{\hat{E}}_{\text{B}_{k}}^{-}(\mathbf{x}_{A},t)\right)|0\rangle^{\text{F}_{k}}|0\rangle^{\text{B}_{k}}\psi_{e}(t)\;\text{for }k\in\left\{ es,ef\right\} \;.\label{eq:psi_k}
\end{eqnarray}
A general solution of the second equation is of the form
\begin{eqnarray*}
\mid\psi_{k}(t)\rangle^{\text{F}_{k},\text{B}_{k}} & = & \frac{i}{\hbar}\int_{t_{0}}^{t}e^{-i\omega_{k}\left(t^{\prime}-t_{0}\right)}\mathbf{d}_{k}\cdot\left(\mathbf{\hat{E}}_{\text{F}_{k}}^{-}(\mathbf{x}_{A},t^{\prime})+\mathbf{\hat{E}}_{\text{B}_{k}}^{-}(\mathbf{x}_{A},t^{\prime})\right)|0\rangle^{\text{F}_{k}}|0\rangle^{\text{B}_{k}}\psi_{e}(t^{\prime})dt^{\prime}\\
 &  & +\mid\psi_{k}(t_{0})\rangle^{\text{F}_{k},\text{B}_{k}}\;\text{for }k\in\left\{ es,ef\right\} \;.
\end{eqnarray*}
Inserting this expression into Eq. (\ref{eq:psi_e}) and using
the initial condition, i.e. 
\[
\mid\psi_{es}(t_{0})\rangle^{\text{F}_{ef},\text{B}_{ef}}=\mid\psi_{\text{in}}\rangle^{\text{F}_{es}}\mid0\rangle^{\text{B}_{ef}}\text{ and }\mid\psi_{ef}(t_{0})\rangle^{\text{F}_{ef},\text{B}_{ef}}=\mid0\rangle^{\text{F}_{ef}}\mid0\rangle^{\text{B}_{ef}},
\]
we obtain the equivalent integro differential equation
\begin{eqnarray*}
\frac{d}{dt}\psi_{e}(t) & = & -\frac{1}{\hbar^{2}}\sum_{k\in\{es,ef\}}\int_{t_{0}}^{t}e^{i\omega_{k}\left(t-t^{\prime}\right)}\langle0|^{\text{F}_{k}}\langle0|^{\text{B}_{k}}\left[\mathbf{d}_{k}^{*}\cdot\left(\mathbf{\hat{E}}_{\text{F}_{k}}^{+}(\mathbf{x}_{A},t)+\mathbf{\hat{E}}_{\text{B}_{k}}^{+}(\mathbf{x}_{A},t)\right)\right]\\
 &  & \left[\mathbf{d}_{k}\cdot\left(\mathbf{\hat{E}}_{\text{F}_{k}}^{-}(\mathbf{x}_{A},t^{\prime})+\mathbf{\hat{E}}_{\text{B}_{k}}^{-}(\mathbf{x}_{A},t^{\prime})\right)\right]|0\rangle^{\text{F}_{k}}|0\rangle^{\text{B}_{k}}\psi_{e}(t^{\prime})dt^{\prime}\\
 &  & +\frac{i}{\hbar}e^{i\omega_{es}\left(t-t_{0}\right)}\langle0|^{\text{F}_{es}}\mathbf{d}_{k}^{*}\cdot\mathbf{\hat{E}}_{\text{F}_{es}}^{+}(\mathbf{x}_{A},t)\mid\psi_{\text{in}}\rangle^{\text{F}_{es}}.
\end{eqnarray*}
\end{widetext}
Within the framework of the above mentioned approximations this expression  yields a complete description of the one photon excitation process of the three level system. In particular, it also describes all possible non Markovian
effects. However, it is well known that in the optical regime with
spontaneous decay rates much smaller than atomic transition frequencies
and for setups in which a photon emitted by an atom does not return
to the atom at a later time, such as in free space or in an open waveguide,
these non Markovian effects are negligible (see e.g.
\cite{mollow1975pure} for an early free space treatment). Hence, in the absence of such photon recurrence phenomena the above
expression simplifies significantly and we obtain
\begin{eqnarray}
\frac{d}{dt}\psi_{e}(t)&=&-\frac{1}{2}\left(\Gamma_{es}+\Gamma_{ef}+\gamma_{es}+\gamma_{ef}\right)\psi_{e}(t)\nonumber\\
&&+i\sqrt{\Gamma_{es}}f_{in}(t)\label{eq:psi_e_simple}
\end{eqnarray}
with
\begin{equation}
\sqrt{\Gamma_{es}}f_{in}(t)=\frac{1}{\hbar}e^{i\omega_{es}\left(t-t_{0}\right)}\langle0\mid\mathbf{d}_{es}^{*}\cdot\left(\mathbf{E}_{es}^{-}\right(\mathbf{x}_{A},t))^{\dagger}\mid\psi_{\text{in}}\rangle^{\text{F}_{es}}\label{eq:f_in_Appendix}
\end{equation}
describing the influence of the incoming single photon wave packet.
The spontaneous
decay rates induced by the reservoir modes $\text{F}_{es}$, $\text{F}_{ef}$,
$\text{B}_{es}$, and $\text{B}_{ef}$
 are denoted by $\Gamma_{es}$,
$\Gamma_{ef}$, $\gamma_{es}$ and $\gamma_{ef}$ . Note that the basic structure
of Eq. (\ref{eq:psi_e_simple}), especially the inhomogeneous term
defined in Eq. (\ref{eq:f_in_Appendix}), closely resembles Eq. (\ref{equationofmotion}) describing the dynamics in the cavity scenario.
In fact, the derivation of both equations follows the same reasoning.}

\bibliographystyle{mau-physrevEng}
\bibliography{References}

\end{document}